\documentclass[onecolumn,usenatbib]{mn2e}
\def\plotfiddle#1#2#3#4#5#6#7{\centering \leavevmode
\vbox to#2{\rule{0pt}{#2}}
\includegraphics{#1}}
\title{Superclusters with thermal SZ effect surveys}
\author[Diaferio et al.]{Antonaldo Diaferio,$^1$ 
 Adi Nusser,$^2$ Naoki Yoshida,$^3$  Rashid A. Sunyaev$^{4,5}$ \\
$^1$Dipartimento di Fisica Generale ``Amedeo Avogadro'', 
Universit\`a degli Studi di Torino, Italy\\
$^2$ Department of Physics, Technion-Israel Institute of Technology, 
Technion City, Haifa, Israel\\
$^3$ Harvard-Smithsonian Center for Astrophysics, Cambridge, MA, USA\\
$^4$ Max-Planck Institut f\"ur Astrophysik, Garching, Germany\\
$^5$ Space Research Institute (IKI), Russian Accademy of Sciences, Moscow, Russia} 
\begin{document}
\maketitle
\begin{abstract}

We use a simple analytic model to compute the angular correlation
function of clusters identified in upcoming thermal 
SZ effect surveys. We then compute the expected fraction of close
pairs of clusters on the sky that are also close along the line of
sight. We show how the expected number of cluster pairs as a function of
redshift is sensitive to the assumed biasing relation between the
cluster and the mass distribution. We find that, in a $\Lambda$CDM model, the fraction 
of physically associated pairs is 
$70\%$ for angular separations smaller than $20$ arcmin and clusters
with specific flux difference larger than $200$ mJy at $143$ GHz.  
The agreement of our analytic results
with the Hubble volume $N$-body simulations is satisfactory.  These
results quantify the feasibility of using SZ surveys to compile
catalogues of superclusters at any redshifts.

\end{abstract}
\begin{keywords}
large scale structure of Universe -- galaxies: clusters: general -- cosmology:
miscellaneous -- methods: analytical 
\end{keywords}

\section{Introduction}

Superclusters are regions with average mass overdensities larger than a few,
on scales larger than a few megaparsecs;  their cluster members 
show indications of  intense 
dynamical activity and provide evidence that structures
form hierarchically on supercluster scales 
(e.g. \citealt*{bardelli01}; \citealt{rines01}; \citealt{plionis02}).
 
Analyses of catalogues of superclusters in the local Universe
(e.g. \citealt{bahcall84}; \citealt{west89}; \citealt{zucca93};
\citealt{einasto01}), beyond the local supercluster \citep{oort83},
have focused on their shapes and their large scale distribution
\citep{kerscher98}. Shape statistics have been used for discriminating
among the cosmological models (\citealt*{basi01}; \citealt*{kolo02}),
whereas the total mass of superclusters can be used to estimate the
density parameter $\Omega_0$ from the mass-to-light ratio on large
scales \citep{small98}. At high redshift, superclusters, when they
exist \citep{postman02}, are difficult to identify, although preferred directions of
radio emission from distant quasars and radio galaxies \citep{west91}
or unusual X-ray morphologies \citep{rosati02} can provide evidence of
their existence at $z\ga 1$.
 
The peculiar velocities of clusters are enhanced in superclusters by
non-linear effects (\citealt{zucca93}; \citealt*{bahcall94}).
\citet{colb00} have indeed shown that the peculiar
velocity of dark matter halos with massive neighbours in $N$-body
simulations is $\sim 40-50\%$ larger than predicted by linear theory.
In fact, linear theory predicts that the evolution of the peculiar
velocities is independent of the local density, whereas \citet{sd01} 
have shown that, in $N$-body simulations, evolved peculiar velocities of halos
with the same mass are larger (smaller) in high (low) density regions.
 Moreover, the high
velocity tail of the cluster peculiar velocity distribution can be a
sensitive discriminator of cosmological models (\citealt{bahcall94}; \citealt{peel02}).
In $N$-body simulations of representative volumes of the Universe, high density
regions of a few megaparsec size contain two or more massive clusters; 
these superclusters can be easily spot in thermal 
\citet{SZ80} effect surveys as nearby CMB temperature decrements in the
Rayleigh-Jeans limit and are the favorite
regions for searches of enhanced kinematic SZ effect \citep*{diaferio00}.
Confusion caused by primary Cosmic Microwave Background (CMB)
anisotropies, thermal SZ and instrumental noise, complicates
the estimation of cluster peculiar velocities from measurements of
 the kinematic effect
\citep*{aghanim01}.  The kinematic SZ effect is difficult to detect
directly; it is roughly an order of magnitude smaller than the thermal
effect. One may proceed by looking for kinematic fluctuations in
clusters selected according  to  thermal SZ measurements.  However the most
massive clusters that are responsible for the largest thermal fluctuations
do not 
necessarily have  the largest peculiar velocities.

The thermal SZ effect, being independent of redshift, 
 is the optimal tool for detecting clusters at high redshifts \citep{carls02}.
Several SZ surveys 
will produce catalogues of thousands clusters at any
redshift in the next few years, e.g. SPT \citep{carls02}, AMI \citep{kneissl01},
AMiBA \citep{lo02}, to mention a few.

Clusters are not randomly distributed on the sky: their angular
correlation function mirrors the clustering evolution of mass.
The angular correlation function of SZ clusters was first computed by
\citet{cole88} who were interested in the non-Gaussianity of 
CMB temperature fluctuations caused by galaxy clusters.
\citet{kom99} show that the correlation in the cluster distribution enhances by $\sim 20-30\%$ 
the Poisson
contribution to the CMB angular power spectrum at degree angular scales.
\citet{mosca02} compute the correlation function on the light-cone
of clusters detectable by the Planck surveyor satellite and show that the correlation length 
of these SZ selected clusters is
more sensitive to the physical properties of the ICM than
to the cosmological parameters.   

In this paper we use a simple analytic model to compute the angular
correlation function and the fraction of pairs that are close along
the line of sight as well as on the sky.  Our result shows that for
viable cosmological models, this fraction is substantial. Therefore if
we define superclusters as regions containing two or more close
clusters, this result motivates the use of thermal SZ cluster surveys
to identify superclusters at any redshift. We also compute the expected number
count of cluster {\it pairs} within a given separation 
 as a function of redshift. Unlike the number count of individual 
clusters, the pair number count is sensitive to the assumed
form of the bias function between the cluster and the mass distribution. 

In section \ref{sec:count} we outline the basic equations; 
in section \ref{sec:stat} we compute the expected number of clusters per solid
angle in flux limited thermal SZ surveys, their
angular correlation function and the number of physically associated
pairs; we then apply our model to a full sky survey like the Planck surveyor. 
In section \ref{sec:bias} we show
how the fraction of physically associated pairs depend on the assumed biasing
relation.  We conclude in section \ref{sec:concl}.

\section{Basic equations and definitions}\label{sec:count}

The interaction between the CMB radiation field and the 
hot cloud of electrons in the ICM
plasma changes the specific intensity of the radiation. The difference 
in the specific intensity 
 $\Delta I_\nu$ at frequency 
$\nu$ is (SZ) 
\begin{equation}
\Delta I_\nu = i_0 g(x) y = i_0 g(x) {k\sigma_T\over m_e c^2}
 \int n_e(r)T_e(r) dl \; ,
\end{equation}
where $i_0=2(kT)^3/(h_Pc)^2$, $k$ is the Boltzmann constant, $T=2.725$ K the present 
day CMB temperature \citep{mather99}, $h_P$ the Planck
constant, $c$ the speed of light, $\sigma_T$ the Thomson cross section, $m_e$
the electron mass, $x=h_P\nu/kT$, 
\begin{equation}
g(x)={x^4 e^x\over (e^x-1)^2} \left(x{e^x+1\over e^x-1}-4\right),
\end{equation}
and $n_e$ and $T_e$ the electron number density and
temperature, respectively. The integral
is over the line of sight $l$, $r^2=w^2+l^2$, where $w$ is the 
separation from the cluster center projected onto the sky, and we assume that 
the cluster is spherically symmetric. 
Here, we neglect the contribution of the kinematic effect.
Let 
\begin{equation}
\langle T_e\rangle_n = {\int n_e(r) T_e(r) dV\over \int n_e(r) dV}
\sim {R_p^2\over N_e^{tot}} \int n_e(r) T_e(r) dl 
\end{equation}
be the electron density weighted mean temperature, where $R_p$ is the projected proper
size of the cluster and $N_e^{tot}$ the total number of electrons in the cluster.
The variation of the CMB specific flux after passing through the cluster is
\begin{equation}
F_\nu  =  \int \Delta I_\nu d\Omega \approx \Delta I_\nu \Delta\Omega
  = i_0 g(x) {k\sigma_T\over m_e c^2}  {f_g\over \mu_e m_p} 
{1\over d_A^2(z)} \langle T_e\rangle_n M   
\end{equation}
where we have assumed an effective solid angle $\Delta\Omega=R_p^2/d_A^2(z)$,
$d_A(z)$ is the angular diameter distance, $N_e^{tot} = M f_g/ \mu_e m_p$,
$M$ is the total mass of the cluster, $f_g$ the gas fraction, $\mu_e$
the mean molecular weight per electron, and $m_p$ the proton mass.
Hereafter, we adopt $f_g=0.06h^{-3/2}$, which has been derived
from a sample of 36 X-ray luminous clusters with redshift in the range $0.05-0.44$
\citep{ef99}; here, 
for simplicity, we neglect (1) the observed variation, up to $\sim 50\%$, 
of $f_g$ between clusters, and (2) the observed dependence of $f_g$ on redshift; 
note that this dependence is weaker 
in low-density Universes than in high-density Universes.
Assuming a fully ionized ICM 
of hydrogen and helium with a helium mass
fraction $Y=0.24$, the mean molecular weight per electron is $\mu_e=(1-Y/2)^{-1}=1.136$.

X-ray observations indicate that the ICM temperature is roughly proportional
to $M^{2/3}$ where $M$ is the virial mass of the cluster (e.g. \citealt*{fin01}; 
\citealt*{al01}), 
in agreement with arguments based on the virial theorem. Hydrodynamical simulations
of cluster formation give  similar  slope for the $T-M$ relation, 
but the temperatures are 
 higher  by  a factor of two or so (e.g. \citealt{mat01}). 
Simulations might miss some important ingredient of the ICM physics;
thus, we adopt a $T-M$ relation in agreement with X-ray observations, i.e.,
\begin{equation}
\langle T_e\rangle_n  = 0.43
[\Delta_c(z)H^2(z)/H_0^2]^{1/3} \left( M\over 10^{14} h^{-1} M_\odot\right)^{2/3}
{\rm keV} \; ,
\label{eq:T-M}
\end{equation}
where $h$ is the present day Hubble constant $H_0$ in units of 100 km s$^{-1}$ Mpc$^{-1}$,
$H^2(z)=H_0^2[\Omega_0(1+z)^3 + (1-\Omega_0-\Lambda_0)(1+z)^2+\Lambda_0]$, and
$\Delta_c(z)$ is the mass density of the cluster, which has just collapsed
at redshift $z$, in units of the critical density at that redshift \citep{bryan}.
Note that here we assume valid the simple top-hat collapse model, although \citet{voit00}
shows that this assumption may be too simplistic when used to derive
the evolution of the mass-temperature relation. 
We also assume that the X-ray temperature equals the electron density
mean weighted temperature $\langle T_e\rangle_n$.

The specific flux variation becomes
\begin{equation}
F_\nu  =  9.33\times 10^4 \left(T\over {\rm K}\right)^3  g(x) {hf_g\over \mu_e }
\left(h^{-1} {\rm Mpc}\over d_A(z)\right)^2  
[\Delta_c(z)H^2(z)/H_0^2]^{1/3} 
\left(M\over 10^{14}h^{-1} M_\odot\right)^{5/3} {\rm mJy}\; .
\end{equation}
SZ surveys will be able to detect only specific flux differences larger
than a threshold $F_\nu^{\rm min}$. This constraint translates into 
a mass threshold $M_{\rm th}$ below which clusters remain undetected:
\begin{equation}
{M_{\rm th}(z,F_\nu^{\rm min})\over 10^{14}h^{-1} M_\odot} = 
1.04 \times 10^{-3} \left(\mu_e \over h f_g\right)^{3/5}
\left(T\over {\rm K}\right)^{-9/5}  
g^{-3/5}(x) \left(d_A(z) \over h^{-1} {\rm Mpc}\right)^{6/5}
[\Delta_c(z)H^2(z)/H_0^2]^{-1/5} 
\left(F_\nu^{\rm min}\over {\rm mJy} \right)^{3/5}.
\end{equation}

\begin{figure}
\plotfiddle{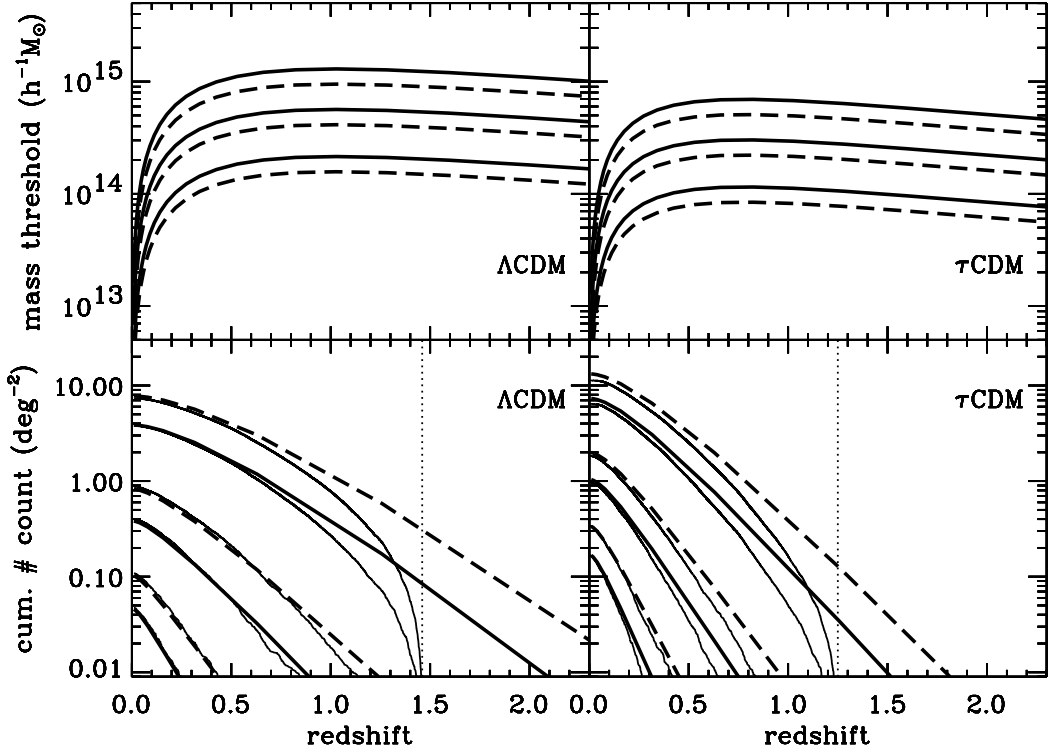}
           {0.4\vsize}              
           {90}                
           {60}                 
           {60}                 
           {230}               
           {-40}                
\caption{Upper panels: mass threshold corresponding to $F^{\rm min}_\nu=200, 50, 10$
mJy (top to bottom) for $\nu=143$ GHz (solid) and $\nu=353$ GHz (dashed lines).
Lower panels: redshift cumulative cluster number counts
corresponding to $F^{\rm min}_\nu=10, 50, 200$
mJy (top to bottom). Bold lines are the expected
distribution (equation \ref{eq:cumul}); thin lines are the actual number from the Hubble volume simulations.
Dotted lines indicate the redshift limit of the lightcone outputs.}
\label{fig:mass_eta}
\end{figure}

The top panels in figure \ref{fig:mass_eta} show $M_{\rm
th}(z,F_\nu^{\rm min})$ for two frequencies measurable by the High
Frequency Instrument (HFI) of the Planck surveyor satellite in two
cosmological models (see section \ref{sec:stat} for details on the
models).  Except for an initial rapid increase, $M_{\rm th}$ varies
slowly with redshift. This feature effectively makes a flux-limited SZ
survey a mass-limited survey \citep{bartlett94}.

\section{Statistics of the cluster distribution}\label{sec:stat}

In this section, we compute 
the redshift distribution of individual clusters, the cluster 
angular correlation function $w(\theta)$, and the redshift
distribution of physically associated cluster pairs.

We compare our analytic calculations  
 with the {\it lightcone} outputs of the two Hubble volume simulations
 (\citealt{evr01}; \citealt{naoki01}). 
These two simulations correspond to  a $\Lambda$CDM and a
 $\tau$CDM cosmology with $(\Omega_0, \Lambda_0, h, \sigma_8)=(0.3,
 0.7, 0.7. 0.9)$ and $(1, 0, 0.5, 0.6)$, respectively.  For both
 models, the power spectrum shape parameter is $\Gamma=0.21$
 (\citealt{frenk00}; see \citealt{evr01} for further details).  Using
 the lightcone outputs has the advantage of simulating a more
 realistic observation of a patch of the sky when compared to the
 usual technique of stacking simulation boxes at different redshifts
 (e.g. \citealt{scaramella93}; \citealt{dasilva00};
 \citealt*{seljak00}; \citealt*{springel01}; \citealt*{zhang02}).
Dark matter halos in the $N$-body simulations have been identified with
a spherical overdensity technique. We assume that these halos
do correspond to galaxy clusters in the appropriate mass range, although
comparisons with the APM cluster catalogue show slightly different clustering properties
\citep{colberg00}.

\subsection{Number count of individual clusters}

The number count $\eta$ per unit solid angle of clusters as microwave
sources, as compared to X-ray sources, was first computed 
by \citet*{korolev86}; many authors, which were interested
in different aspects of $\eta$, have computed it ever since 
(see e.g. \citealt{bart01} and
references therein). The number count $\eta$ enters the calculation of the angular
correlation function (section \ref{sec:wtheta}). It 
is sensitive to the properties of the ICM
(\citealt{hc01}; \citealt{benson02}) and to the cosmological
parameters (\citealt{barbosa96}; \citealt{kneissl01}; \citealt*{haiman01}; 
\citealt*{holder01}; \citealt*{benson02}). 
 The number count also depends on the parameters of the
survey, namely its sensitivity (or its flux limit $F_\nu^{\rm lim}$)
and its angular resolution.  The cumulative number
count of clusters, at redshift greater than $z$, detected per unit
solid angle is 
\begin{equation}
\eta(z,F_\nu^{\rm min})d\Omega  = d\Omega\; 
c\int_z^\infty {d^2(z_1) \over H(z_1)} dz_1
\int_{M_{\rm th}(z_1,F_\nu^{\rm min})}^\infty {dn(M,z_1)\over dM} dM
\label{eq:cumul}
\end{equation}
where $d(z)=(1+z)d_A(z)$ is the distance measure, and  
$dn(M,z)/dM$ is the comoving number density of clusters with mass in the range $(M,M+dM)$
at redshift $z$.
We have assumed that the detector is able to measure the total specific
flux difference $F_\nu$ of each cluster. When the
angular resolution of the survey is included in the
analysis, the cluster number count can drop significantly \citep{bart00}.  

In the Hubble volume simulations, the dark matter halo mass function is well
fitted by the formula \citep{jenk01}
\begin{equation}
{dn_{J}(M,z)\over dM}dM={A\langle\rho\rangle\over M}\alpha_{\rm eff}(M)\exp[-\vert\ln\sigma^{-1}(M)+B\vert^\epsilon] dM
\label{eq:jmf}
\end{equation}
where $(A, B, \epsilon)=(0.22, 0.73, 3.86)$ and $(0.27, 0.65, 3.77)$
for the $\Lambda$CDM and $\tau$CDM model, respectively,
$\langle\rho\rangle$ is the comoving mean density,
$\alpha_{\rm eff}=d\ln\sigma^{-1}(M)/d\ln M$,
\begin{equation}
\ln\sigma^{-1}(M)=-\ln\sigma_{15}+a\ln M+b(\ln M)^2
\end{equation}
and $(\sigma_{15}, a, b)=(0.578, 0.281, 0.0123)$ and $(0.527, 0.267, 0.0122)$
for the $\Lambda$CDM and $\tau$CDM model, respectively.

The lower panels of figure \ref{fig:mass_eta}
show the redshift cumulative cluster number count for the two
models (thin lines) compared with the expectation of equation (\ref{eq:cumul}).
 We use the {\it NO} lightcone outputs of the Hubble volume 
simulations, which cover
$\pi/2$ sr and have depth
$z=1.46$ ($\Lambda$CDM) and $1.25$ ($\tau$CDM), as indicated by the dotted lines.
The disagreement at low $F^{\rm min}_\nu$ and at high redshift is mainly due to
the mass resolution limit.  

\subsection{The angular correlation function}\label{sec:wtheta}

 To compute $w(\theta)$, we
need to know the three dimensional two-point correlation function
$\xi(r_{12},M_1,M_2,z_1,z_2)$ for clusters of mass $M_1$, $M_2$ at
redshift $z_1$, $z_2$ and comoving separation $r_{12}$. When
$z_1=z_2=z$, we can use the simple biasing scheme
\begin{equation}
\xi(r_{12},M_1,M_2,z)=b_{\rm ST}(M_1,z)b_{\rm ST}(M_2,z)\xi_m(r_{12},z)
\label{eq:xi12}
\end{equation}
where \citep{st99}
\begin{equation}
b_{\rm ST}(M,z)=1+{a\nu^2-1\over \delta_c(z)} + {2p\over \delta_c(z) [1+ (a\nu^2)^p]},
\label{eq:bias_ST}
\end{equation}
$a=0.707$, $p=0.3$, $\nu=\delta_c(z)/\sigma(M,z)$,
$\sigma(M,z)$ is the rms of the mass density field smoothed on scale $R=
(3M/4\pi \langle\rho\rangle)^{1/3}$, $\langle\rho\rangle$ is the mean comoving
mass density, and $\delta_c(z)$ is the linear overdensity extrapolated at
redshift $z$ for a spherical perturbation which has just collapsed (see 
\citealt{kitayama96} or \citealt{nakamura97} 
for handy fitting formulae to these quantities);  
$\xi_m(r,z)$ is the correlation function of dark matter at redshift $z$.
Equation (\ref{eq:bias_ST}) is strictly valid on large scales. Because we are interested
in close pairs of clusters, we need to consider a scale dependent bias factor
$b(r,M,z)$ (e. g. \citealt{yoshi01}). 
For clusters at different redshifts $z_1$ and $z_2$,
we can modify equation (\ref{eq:xi12}) to
\begin{equation}
\xi(r_{12},M_1,M_2,z_1,z_2)=
b(r_{12},M_1,z_1)b(r_{12},M_2,z_2)\xi_m(r_{12},z_1)
\label{eq:xiz12}
\end{equation}
where we break the symmetry in $z_1$ and $z_2$ by ``sitting'' on the cluster
at $z_1$. Equation (\ref{eq:xiz12}) is consistent with equation (\ref{eq:xi12}), because
clusters at different epochs $z_1$, $z_2$ are not correlated and
$\xi(r_{12})$ goes to zero with increasing $r_{12}$ quickly enough that
$\xi(r_{12})\ne 0$ only when $z_1\sim z_2$. 
Alternative choices of the redshift at which $\xi_m$ 
can be computed are discussed in \citet{porciani97}.

The comoving relative separation $r_{12}$ in a space described by the 
Robertson-Walker metric is (\citealt{weinberg72}, equation 14.2.7; \citealt{osm81}) 
\begin{equation}
r_{12}^2(z_1,z_2,\theta) =
d^2(z_1) G^2(z_1,z_2,\theta) + d^2(z_2) -  2 d(z_1)d(z_2)G(z_1,z_2,\theta) \cos\theta
\end{equation}
where $\theta$ is the angular separation of the two objects on the sky, 
\begin{equation}
G(z_1,z_2,\theta)  =
\sqrt{1-\kappa d^2(z_2)} + \left(1-\sqrt{1-\kappa d^2(z_1)}\right)
 {d(z_2)\over d(z_1)} \cos\theta
\end{equation} 
and $\kappa = (H_0/ c)^2 (\Omega_0+\Lambda_0-1)$. This equation is {\it not}
symmetric in $z_1$ and $z_2$ for a non flat Universe ($\kappa \ne 0$). 
However, if we deal with distances
smaller than the curvature radius of the Universe $1/\sqrt{\vert\kappa\vert}$, we have
$1/\kappa \gg d^2(z_1)$ and $\gg d^2(z_2)$, and $G\sim 1$; this is the
case of interest, because we expect $\xi(r_{12})\sim 0$ at large separation.
When $z_1=z_2=z$, $r_{12}^2\sim 2d^2(1-\cos\theta)$
for $\kappa d^2\sim 0$, and we correctly recover
$\theta=r_{12}/d(z)$ for small $\theta$.

To compute the angular correlation function, we follow the classical argument that
leads to Limber's equation (\citealt{mat97}; \citealt{mosca00}). 
Consider the probability of finding two clusters 
with different mass at different redshift
\begin{equation}
\zeta(r_{12},M_1,M_2,z_1,z_2)dA_1dA_2=
 dV_1dV_2 {dn(M_1,z_1)\over dM_1}dM_1{dn(M_2,z_2)\over dM_2}dM_2
 [1+\xi(r_{12},M_1,M_2,z_1,z_2)]
\end{equation}
where $dA_i=dM_idz_id\Omega_i$, $d\Omega_i$ is the infinitesimal solid angle
 and $dV_i=c d^2(z_i)d\Omega_idz_i/H(z_i)$ is 
the infinitesimal comoving volume. The probability of detecting two clusters
at angular separation $\theta$ is proportional to
\begin{eqnarray}
\psi(\theta)d\Omega_1d\Omega_2 & = & d\Omega_1d\Omega_2 \int 
\zeta(r_{12},M_1,M_2,z_1,z_2) dM_1dM_2dz_1dz_2\cr
& = & d\Omega_1d\Omega_2\eta_0^2(F_\nu^{\rm min})\left[1 + w(\theta,F_\nu^{\rm min})\right] 
\label{eq:psi}
\end{eqnarray}
where $\eta_0(F_\nu^{\rm min})=\eta(0,F_\nu^{\rm min})$, and 
\begin{equation}
w(\theta,F_\nu^{\rm min}) = {\Xi_0(\theta,F_\nu^{\rm min})\over \eta_0^2(F_\nu^{\rm min})}
\label{eq:wtheta}
\end{equation}
is the angular correlation function we are seeking.
All the clustering and biasing information is in the function 
\begin{eqnarray}
\Xi(z,\theta,F_\nu^{\rm min})&=&c^2\int_z^\infty 
n(z_1,F_\nu^{\rm min}) {d^2(z_1)\over H(z_1)} dz_1 \times \cr
&\phantom{=}& \times \int_0^\infty n(z_2,F_\nu^{\rm min}) 
b_{\rm eff}(r_{12},z_1,F_\nu^{\rm min}) 
b_{\rm eff}(r_{12},z_2,F_\nu^{\rm min}) \xi_m(r_{12},z_1)
{d^2(z_2)\over H(z_2)}dz_2
\label{eq:Xi}
\end{eqnarray}
where
\begin{equation}
n(z,F_\nu^{\rm min})=\int_{M_{\rm th}(z,F_\nu^{\rm min})}^\infty {dn(M,z)\over dM}dM
\end{equation}
\begin{equation}
b_{\rm eff}(r,z,F_\nu^{\rm min})=n^{-1}(z,F_\nu^{\rm min})
\int_{M_{\rm th}(z,F_\nu^{\rm min})}^\infty b(r,M,z) {dn(M,z)\over dM}dM
\label{eq:beff}
\end{equation}
and $\Xi_0(\theta,F_\nu^{\rm min})= \Xi(0,\theta,F_\nu^{\rm min})$.

\begin{figure}
\plotfiddle{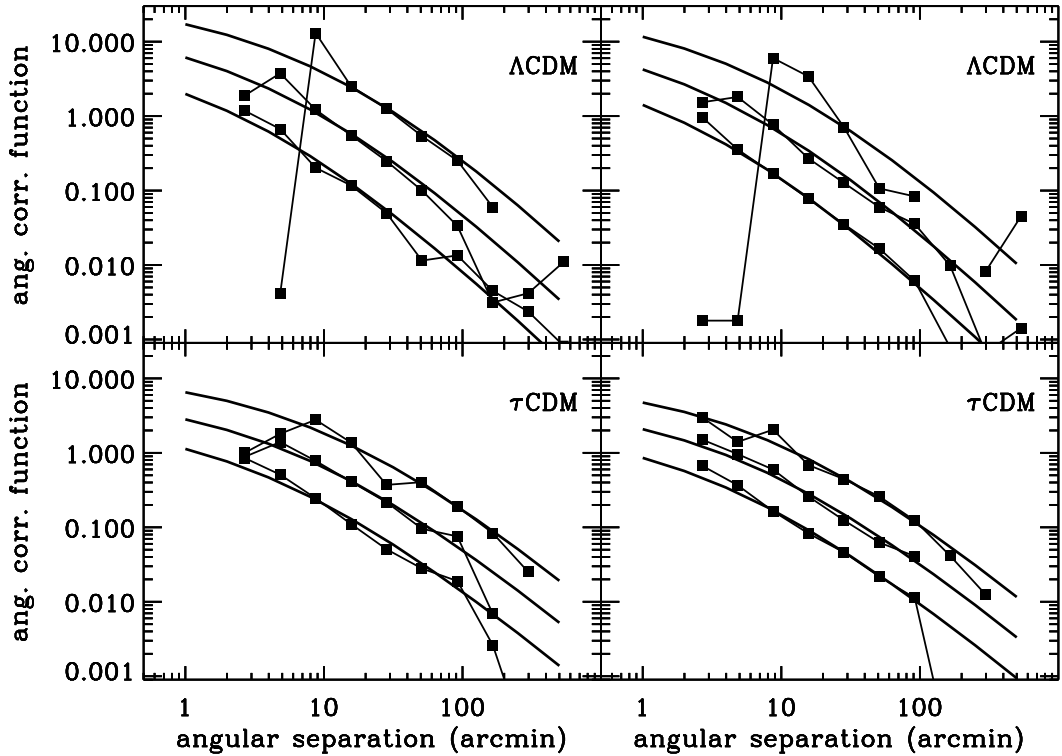}
           {0.4\vsize}              
           {90}                
           {60}                 
           {60}                 
           {230}               
           {-40}                
\caption{Angular correlation function of SZ clusters with $F^{\rm min}_\nu>200, 50, 10$
mJy (top to bottom) for $\nu=143$ GHz (left panels) and $\nu=353$ GHz (right panels).
Bold lines are the expected angular correlation function (equation \ref{eq:wtheta}); 
filled squares
are the correlation function from the Hubble volume lightcones.}
\label{fig:wtheta}
\end{figure}

We use the fitting formula given by \citet{PD}
to compute the correlation function of dark matter $\xi_m(r_{12},z)$ in the non-linear regime.
For the effective scale dependent bias factor $b_{\rm eff}$ 
we adopt the following fitting formula.
\begin{equation}
b_{\rm eff}(r,z,F_\nu^{\rm min})=b_{\rm ST, eff}(z,F_\nu^{\rm min})
[1+b_{\rm ST, eff}(z,F_\nu^{\rm min})\sigma(r,z)]^{0.35} \; ,
\label{eq:bias}
\end{equation}
where $\sigma(r,z)$ is the mass variance smoothed over the top-hat radius $r$, and
\begin{equation}
b_{\rm ST, eff}(z,F_\nu^{\rm min})=
{\int_{M_{\rm th}(z,F_\nu^{\rm min})}^\infty 
b_{\rm ST}(M,z) [dn_J(M,z)/ dM]dM\over
\int_{M_{\rm th}(z,F_\nu^{\rm min})}^\infty [dn_J(M,z)/ dM]dM } \; .
\end{equation}
Figure \ref{fig:wtheta} shows that this choice of the biasing relation yields
an excellent agreement between the model and the angular correlation function 
derived from the pair count of SZ clusters in the $N$-body simulations.
The functional form of the biasing relation (equation \ref{eq:bias}) 
was first suggested by \citet{hamana02},
with a power 0.15 rather than 0.35. Hamana et al's relation was only
checked against the redshift zero output. When we compare the analytic modelling
with the lightcone output, we include a broad redshift range, and the agreement
shown in figure \ref{fig:wtheta} indicates that a larger power seems more
appropriate. 

\subsection{The number of physically associated cluster pairs}\label{sec:pairs}

The fraction of pairs with angular separation smaller than $\theta$
whose cluster members have comoving separation along the line of sight
smaller than $r_{\rm max}$ is proportional to an expression similar to
equation (\ref{eq:Xi}) where the integral over $z_2$ is limited to the
interval set by $r_{\rm max}$.  Because $\xi(r_{12})\sim 0$ at large
$r_{12}$, when $r_{\rm max}$ is of order the cluster correlation
length, we can estimate this fraction as
\begin{equation}
\varphi_0(<\theta) = {\int_0^\theta w(\theta_1)\theta_1d\theta_1\over \int_0^\theta[1+w(\theta_1)]
\theta_1d\theta_1}.
\label{eq:varphi_0}
\end{equation} 
Thus, $\varphi_0(<\theta)$ also indicates the fraction of close pairs
of clusters which are roughly at the same redshift.

\begin{figure}
\plotfiddle{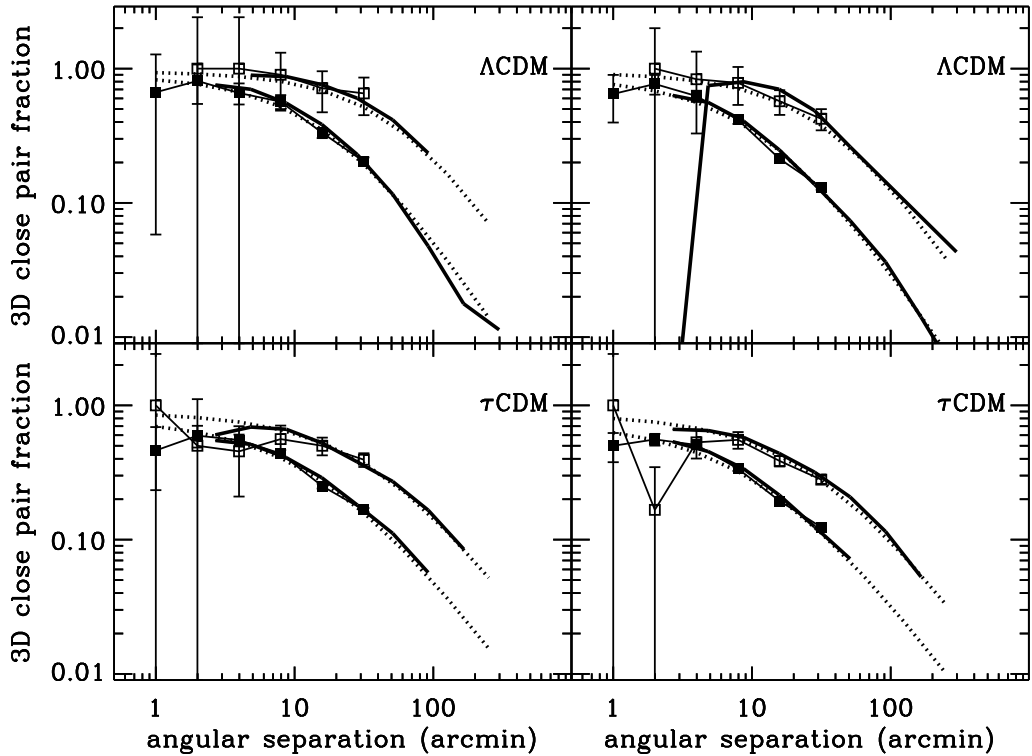}
           {0.9\vsize}              
           {90}                
           {60}                 
           {60}                 
           {230}               
           {-40}                
\caption{Fraction of cluster pairs with comoving separation $<16.0$ ($10.7$) $h^{-1}$ Mpc 
for the $\Lambda$CDM 
($\tau$CDM) model (squares). Upper (lower) curves are for $F_\nu^{\rm min}>200$ $(50)$ mJy.
Left (right) panels are for $\nu=143$ ($353$) GHz.
Bold lines correspond to $\varphi_0(<\theta)$ where $w(\theta)$ is computed
directly from the $N$-body simulations. Dotted lines are the corresponding
analytic functions (equation \ref{eq:varphi_0}). Error bars assume Poisson statistics.}
\label{fig:pairs}
\end{figure} 

When comparing $\varphi_0(<\theta)$ with the actual pair count in 
the $N$-body simulations, we need to quantify what
we mean by {\it close} along the line of sight.  We compute
$\varphi_0(<\theta)$ with $w(\theta)$ derived from the simulations and
we choose the maximum comoving pair separation that provides a
reasonable agreement between $\varphi_0(<\theta)$ and the actual count
of close pairs.  The comoving volume of the $\tau$CDM model is
$(2/3)^3$ smaller than the comoving volume of the $\Lambda$CDM model,
but the two volumes contain the same total mass; we thus choose, for
the two models, two different maximum comoving separations whose ratio
is $2/3$: therefore, we have the same mass within the volumes defined
by these separations.  Figure \ref{fig:pairs} shows that the agreement between 
$\varphi_0(<\theta)$ derived from the simulations (bold lines) and the
actual pair count (squares) with cluster member three-dimensional
comoving separation smaller than 16 (10.7) $h^{-1}$ Mpc for the $\Lambda$CDM
($\tau$CDM) model is satisfactory.  Dotted lines show $\varphi_0(<\theta)$
computed with the analytic model. 
The agreement between the simulations and the analytic model is excellent.

Figure \ref{fig:pairs} shows that the analytic model predicts that
more than $\sim 70\%$ of the pairs with angular separation $\theta
<20$ arcmin are physical associations in the shallow survey
$F_\nu^{\rm min}>200$ mJy at $\nu=143$ GHz.  Because of the larger number of
clusters per square degree, the fraction of pairs close in three
dimensions drops to $\sim 30\%$ in the deeper survey $F_\nu^{\rm
min}>50$ mJy; this fraction is still substantial.  \citet{mosca02}
show that, because the properties of the ICM affect the cluster mass
threshold, for a fixed specific flux threshold $F_\nu^{\rm min}$, the
correlation length $r_0$ measured for SZ clusters is more sensitive to
the ICM properties than to the cosmological parameters.  According to
their results, however, our estimated fractions of physically
associated pairs can change by 30\% at most.

\begin{figure}
\plotfiddle{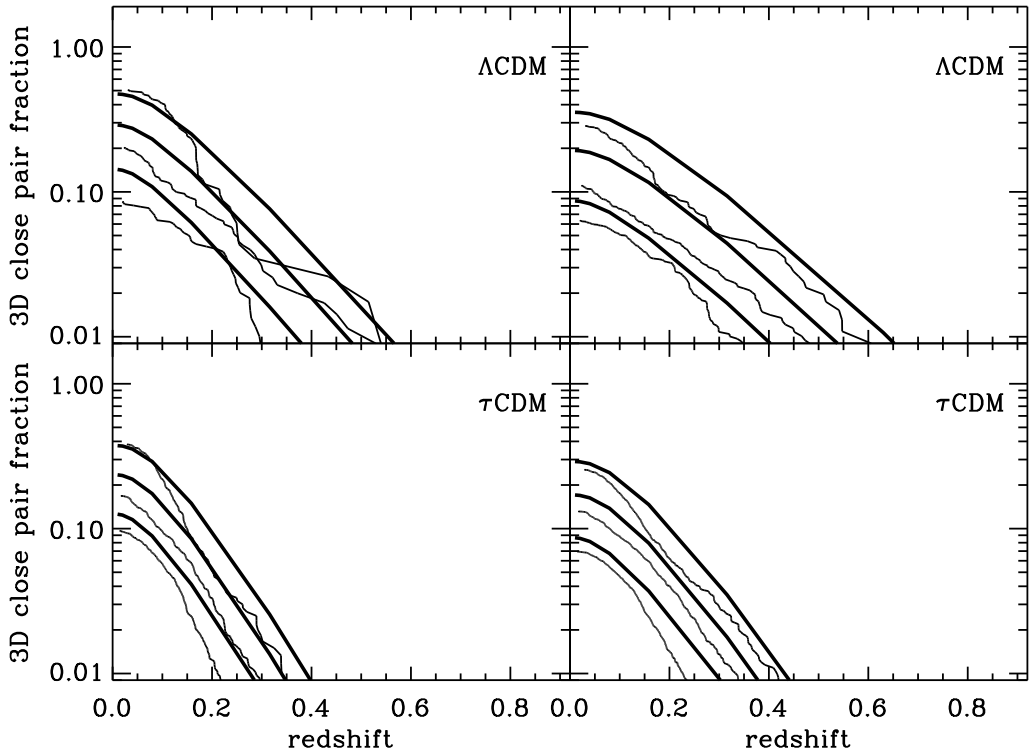}
           {0.9\vsize}              
           {90}                
           {60}                 
           {60}                 
           {230}               
           {-40}                
\caption{Cumulative redshift distribution of the fraction of cluster 
pairs with comoving separation $<16.0$ ($10.7$) $h^{-1}$ Mpc, $F_\nu^{\rm min}> 50$ mJy,
and angular separation in the range $(8,16), (16,32)$ and $(32,64)$ 
arcmin (top to bottom) for the $\Lambda$CDM
($\tau$CDM) $N$-body models (thin lines). 
Left (right) panels are for $\nu=143$ ($353$) GHz.
Bold lines show the corresponding analytic distributions $\varphi(z,\theta_1,\theta_2)$ 
(equation \ref{eq:cumul_z}).}
\label{fig:pairs_z}
\end{figure} 

The cumulative redshift distribution of the fraction of physically associated cluster pairs 
with angular separation in the range ($\theta_1$, $\theta_2$) is 
\begin{equation}
\varphi(z,\theta_1,\theta_2) = {\int_{\theta_1}^{\theta_2}
\Xi(z,\theta,F_\nu^{\rm min})\theta d\theta \over 
\int_{\theta_1}^{\theta_2} [\eta_0^2(F_\nu^{\rm min}) + \Xi_0(\theta,F_\nu^{\rm min})
] \theta d\theta}.
\label{eq:cumul_z}
\end{equation}
Figure \ref{fig:pairs_z} compares the model with the simulations. 
The agreement is satisfactory.
Shallower surveys will detect only massive clusters
which form at late times. Therefore most pairs of close clusters are
at moderately low redshift.

To compute the number of pairs with separation $\theta$ on the sky, consider
 the probability $\eta_0^2[1+w(\theta)]d\Omega_1 d\Omega_2$ 
of having such a pair, where $d\Omega_i$ is the
infinitesimal solid angle around each pair member and the angle
between the versors pointing to $d\Omega_1$ and $d\Omega_2$ is
$\theta$. Without loss of generality, we can fix the versor of $d\Omega_1$
coincident with the $z$-axis and write $d\Omega_2=d\phi_2
d\cos\theta_2\equiv d\phi_2 d\cos\theta$.  Integrating over
$d\Omega_1d\phi_2$ yields $\int d\Omega_1d\Omega_2= 8\pi\sin\theta
d\theta$.  All over the sky, the number of pairs with separation in
the range $(\theta,\theta+d\theta)$ is thus $\tilde n_p(\theta)
d\theta =8\pi^2\eta_0^2[1+w(\theta)]\sin\theta d\theta/ (2+\langle
w\rangle)$, where $\langle w\rangle=\int_0^\pi \sin\theta
w(\theta)d\theta$ and the relation is normalized such that it yields a
total of $(4\pi\eta_0)^2/2$ pairs at any separation. In a survey with
area $\Delta \Omega$, the number of pairs with angular separation in
the range $(\theta_1,\theta_2)$ is trivially
$(\Delta\Omega/4\pi)\int_{\theta_1}^{\theta_2} \tilde n_p(\theta)
d\theta$. We use this relation in the next section.

\subsection{A survey example}\label{sec:example}

\begin{figure}
\plotfiddle{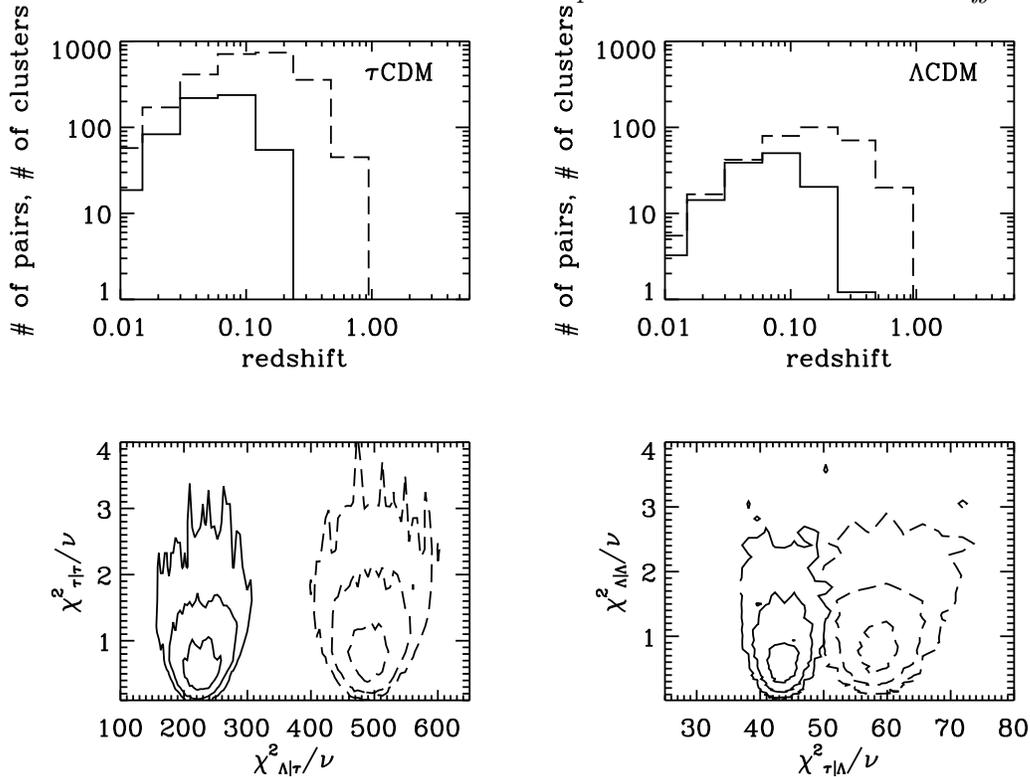}
           {0.8\vsize}              
           {90}                
           {60}                 
           {60}                 
           {230}               
           {-40}                
\caption{Upper panels: differential redshift distribution $n_p(z)$ 
of physically associated pairs of clusters with angular separation in the range
8--16 arcmin for a full sky survey with $F_\nu^{\rm min}= 100$ mJy
and $\nu=143$ GHz (solid histograms); differential redshift distribution $n_c(z)$ 
of all the clusters in pairs with angular separation in the range 8--16 arcmin, 
irrespective of their physical association, in the same survey (dashed histograms). 
Lower panels: probability density of the reduced $\chi^2$ of Monte Carlo realizations 
drawn
from the $\tau$CDM model (left panel) and the $\Lambda$CDM model (right panel)
computed with respect to each model. Solid (dashed) contours are for the pair
(cluster) samples. Contour levels are (0.61,0.13,0.01) times the maximum
probability density. They roughly correspond to the 68\%, 95\% and 99\% 
confidence levels.    }
\label{fig:chi2}
\end{figure}

To get a better idea of the sample sizes,
consider a full sky survey, like the one that will be performed,
for example, by the Planck surveyor, with a
specific flux difference limit 100 mJy at 143 GHz.  In a $\Lambda$CDM
($\tau$CDM) model, this survey would detected $\sim 5500$ (17400)
clusters and $\sim 155$ (1150) pairs with angular separation in the
range 8--16 arcmin. Of these pairs, $83\%$ (53\%) are also close along the line
of sight.  It means that measuring the
redshift of $155\times 2$ ($1150\times 2$) clusters in close pairs
would yield $128$ ($614$) physically associated pairs and $54$
($1052$) ``isolated'' clusters.  These numbers are computed with the
analytic model described above.  The differential
redshift distributions $\bar n_c(z)$ of the $310$ ($2300$) clusters and
$\bar n_p(z)$ of the 128 (614) pairs are shown in figure \ref{fig:chi2}.

Extracting the cosmology information from the cluster redshift distribution is
not trivial.  The redshift distribution of clusters depends on the
cosmological parameters in a
complicated way, through the geometry of the Universe and the cluster
mass function (equation \ref{eq:cumul}).  The redshift distribution of
pairs introduces further complications due to the cluster bias
function and the non linear evolution of the dark matter correlation
function (equation \ref{eq:Xi}). So an estimation of the cosmological
parameters with a traditional least squares fitting procedure is very
cumbersome.  A simpler procedure is to compare the observed
distributions with those predicted by a few fiducial
models. Therefore, we perform the following Monte Carlo
simulation. First, by assuming a cosmological model, we use our analytical model to compute 
the expected number of clusters $\bar n_c(z)$ and pairs
of clusters $\bar n_p(z)$ in a given redshift bin. We then generate samples
of clusters and cluster pairs from Poisson distributions with mean
values $\bar n_c(z)$ and $\bar n_p(z)$, respectively.

Finally we compute 
\begin{equation}
\chi^2_{i\vert j} = \sum_{z {\rm bin}} {(n_j-\bar n_i)^2
\over \bar n_i}
\label{eq:chi2}
\end{equation}
where the sum is over the redshift bins, the indices $i$ and $j$ refer
to the cosmological model (e.g., $\Lambda$CDM or $\tau$CDM), $n_j$ is
the variable drawn from model $j$ and $\bar n_i$ is the expected mean
value in model $i$.  

The lower panels of figure \ref{fig:chi2} show the distribution of the
$\chi^2$ for $10^4$ Monte Carlo realizations drawn from the redshift
distributions shown in the upper panels.  Each random sample (either
of clusters or pairs of clusters) drawn from a model $j$ is a point on
the plane $\chi^2_{\tau\vert j}-\chi^2_{\Lambda\vert j}$ in
 figure \ref{fig:chi2}.  Both the
cluster and pair samples would give significant large $\chi^2$ when
compared to the wrong cosmological model. 

\subsection{Sensitivity on the biasing relation}\label{sec:bias}

The predicted number of physically associated cluster pairs depends on the assumed biasing
relation between the cluster and the mass distribution.
In the previous sections, we have assumed a time and scale dependent 
bias factor (equation \ref{eq:bias}).
To investigate how  sensitive the fraction of pairs is on this assumption, we consider two 
extreme models: a scale independent and a time independent bias factors. Namely,
\begin{equation}
b_{{\rm no}-r}(z,F_\nu^{\rm min})=b_{\rm ST, eff}(z,F_\nu^{\rm min}) 
\label{eq:bias_no-r}
\end{equation}
and
\begin{equation}
b_{{\rm no}-z}(r,F_\nu^{\rm min})=b_{\rm ST, eff}(0,F_\nu^{\rm min})
[1+b_{\rm ST, eff}(0,F_\nu^{\rm min})\sigma(r,0)]^{0.35} \; .
\label{eq:bias_no-z}
\end{equation}

\begin{figure}
\plotfiddle{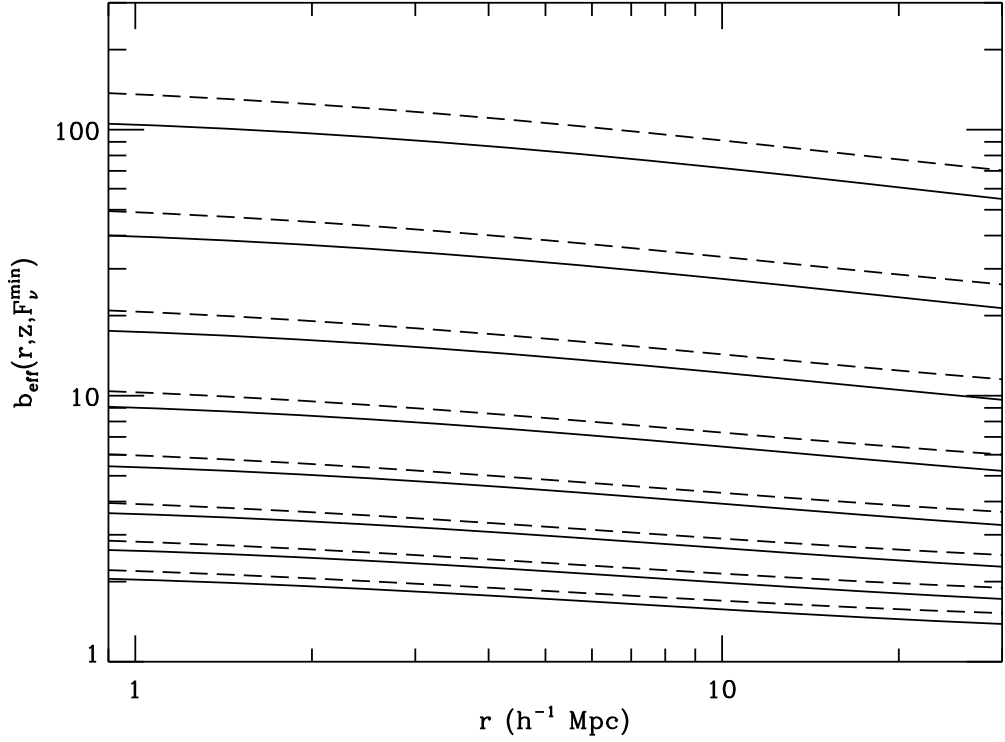}
           {0.9\vsize}              
           {90}                
           {60}                 
           {60}                 
           {230}               
           {-40}                
\caption{Effective bias (equation \ref{eq:bias}) of SZ selected clusters in a
survey with $F_\nu^{\rm min}= 50$ mJy
and $\nu=143$ GHz. Solid (dashed) lines are for the $\Lambda$CDM ($\tau$CDM) model. 
Curves are for $z=0.01, 0.02, 0.06, 0.14, 0.34, 0.82, 2.0, 4.8$ (bottom to top).}
\label{fig:bias_r}
\end{figure}   

\begin{figure}
\plotfiddle{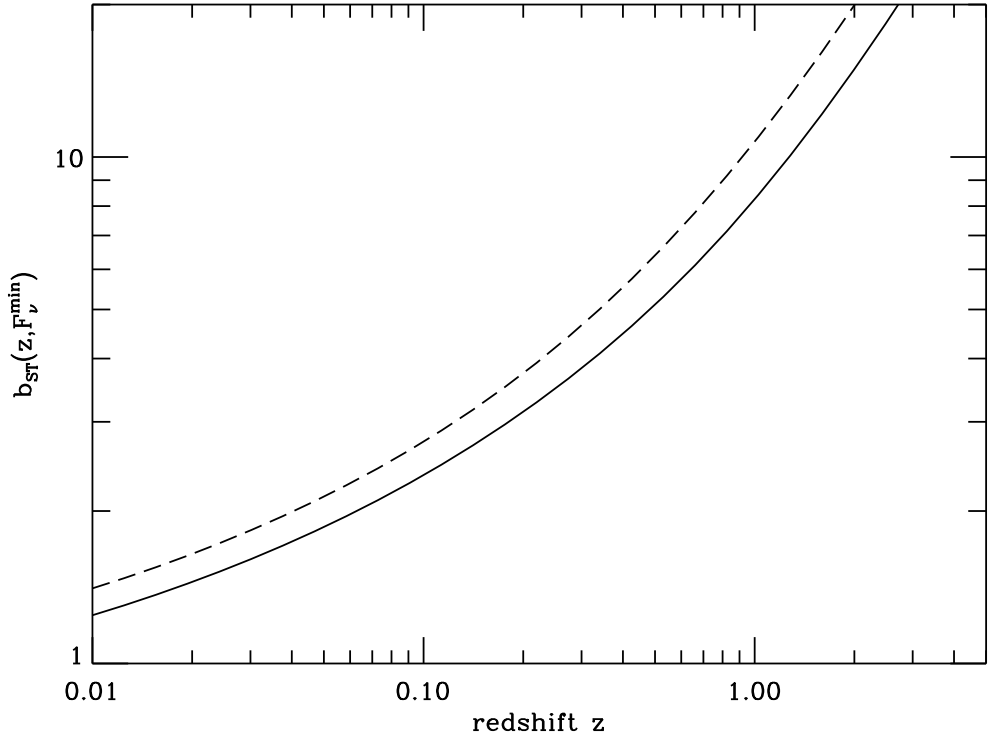}
           {0.9\vsize}              
           {90}                
           {60}                 
           {60}                 
           {230}               
           {-40}                
\caption{Effective scale independent bias (equation \ref{eq:bias_no-r}) 
of SZ selected clusters in a 
survey with $F_\nu^{\rm min}= 50$ mJy
and $\nu=143$ GHz. Solid (dashed) line is for the $\Lambda$CDM ($\tau$CDM) model.  }
\label{fig:bias_st}
\end{figure} 

Figure \ref{fig:bias_r} shows our fiducial biasing relation 
$b_{{\rm eff}}(r,z,F_\nu^{\rm min})$;
in this figure, the bottom lines of the $\Lambda$CDM and $\tau$CDM models show 
the time independent bias factor $b_{{\rm no}-z}(r,F_\nu^{\rm min})$. 
Figure \ref{fig:bias_st} show the scale independent bias factor 
$b_{{\rm no}-r}(z,F_\nu^{\rm min})$. 

\begin{figure}
\plotfiddle{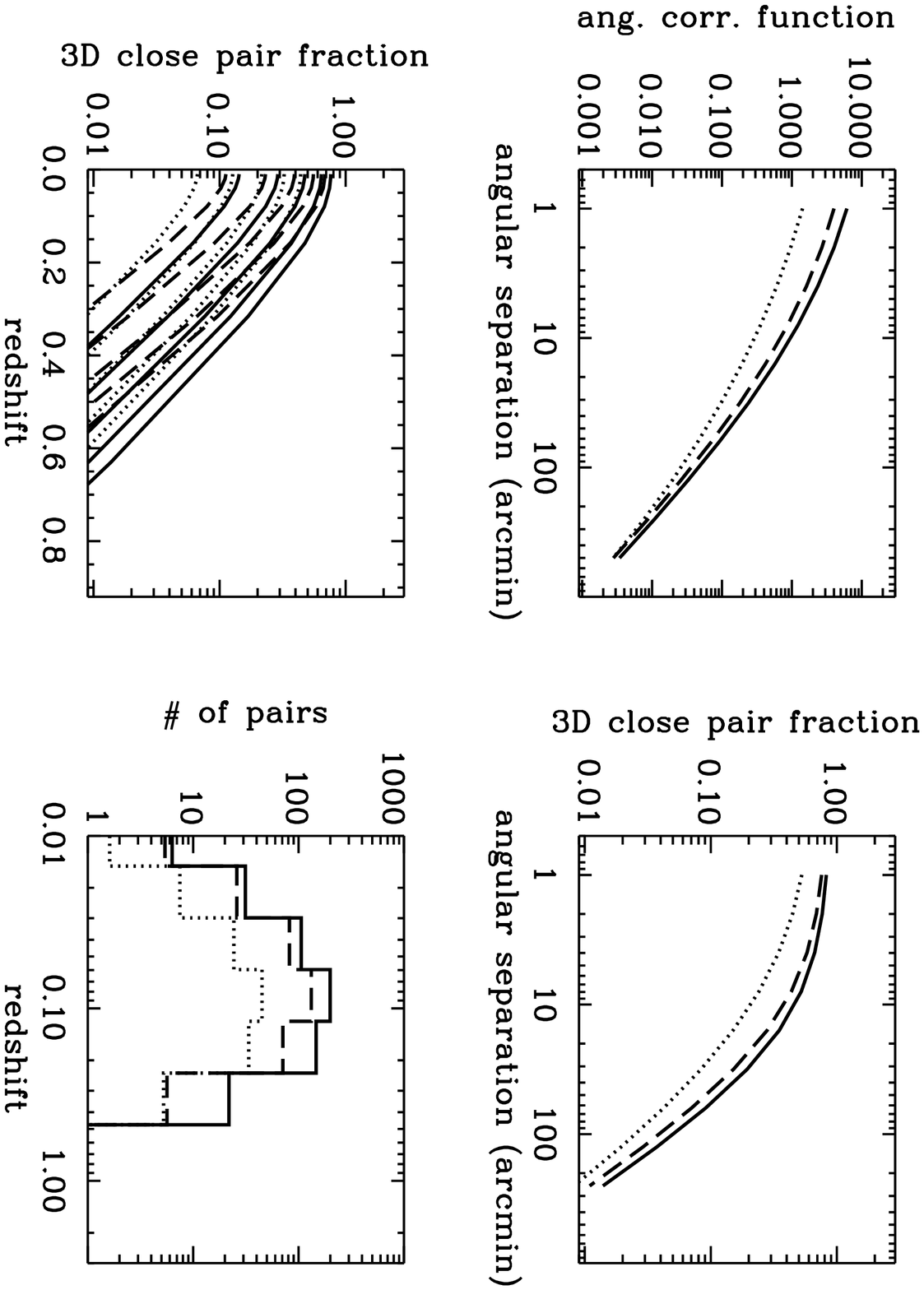}
           {0.9\vsize}              
           {90}                
           {60}                 
           {60}                 
           {230}               
           {-40}                
\caption{Sensitivity of our results on the biasing relation in the $\Lambda$CDM model
for a survey with $F_\nu^{\rm min}= 50$ mJy and $\nu=143$ GHz. 
Solid, dashed and dotted lines refer to 
our fiducial (equation \ref{eq:bias}), time independent (equation \ref{eq:bias_no-z}) 
and scale independent
(equation \ref{eq:bias_no-r}) biasing relations, respectively. 
The bottom right panel shows the redshift distribution of physically associated
cluster pairs with angular separation in the range 8-16 arcmin in a full sky
survey.}
\label{fig:bias_comp}
\end{figure}

Figure \ref{fig:bias_comp} shows that neglecting the scale 
dependence considerably suppresses the angular correlation
function (dotted lines). Because clusters form
at relatively low redshift, neglecting the time dependence is less dramatic
(dashed lines): in fact,
the number of physically associated pairs is substantially underestimated only at
$z>0.2$ (bottom panels).

\begin{figure}
\plotfiddle{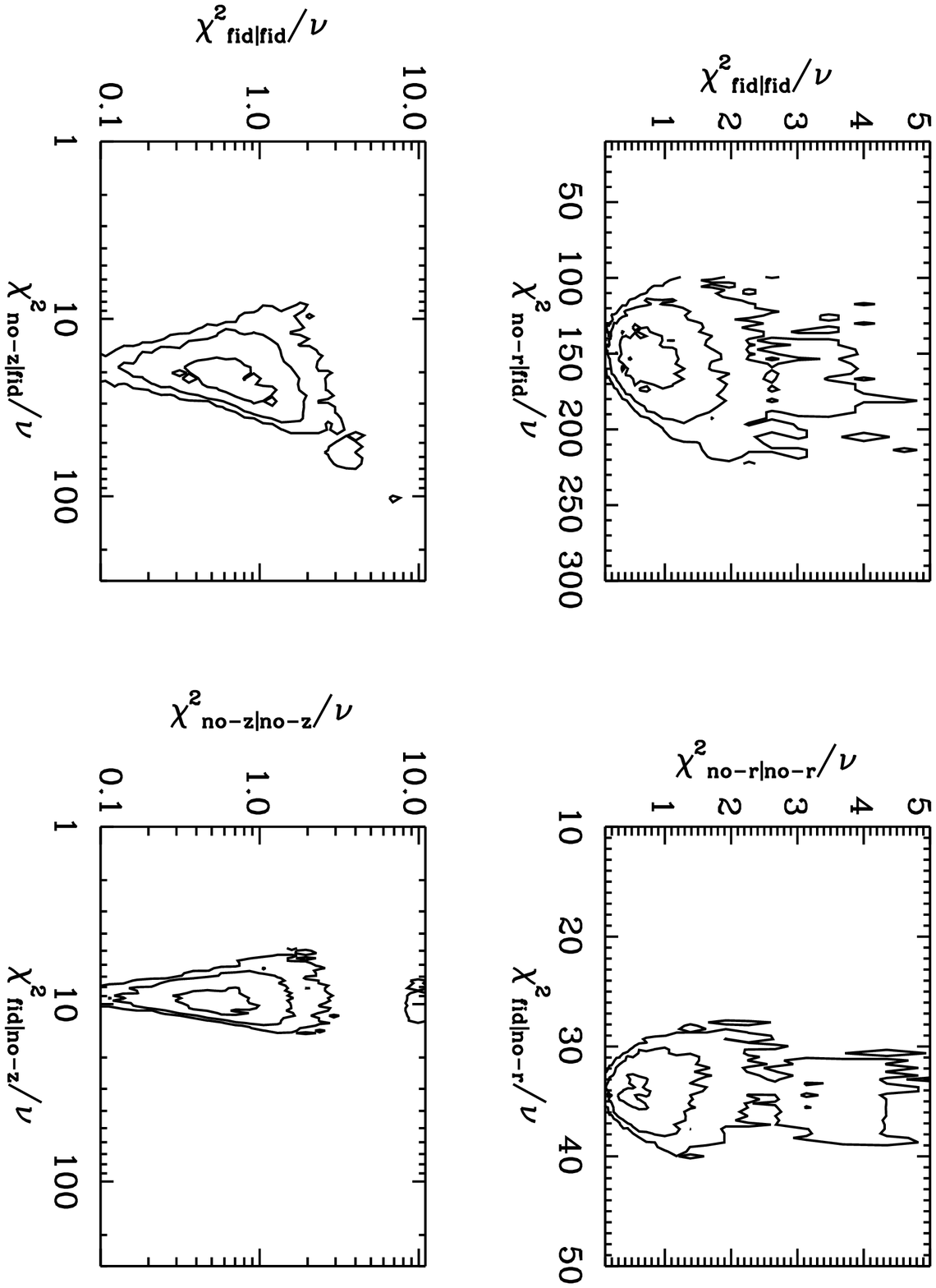}
           {0.9\vsize}              
           {90}                
           {60}                 
           {60}                 
           {230}               
           {-40}                
\caption{Values of the $\chi^2$ corresponding to the bottom right panel of
figure \ref{fig:bias_comp}. See text for more information.
Contour levels are (0.61,0.13,0.01) times the maximum
probability density. They roughly correspond to the 68\%, 95\% and 99\%
confidence levels.} 
\label{fig:bias_chi2}
\end{figure}  
             
To quantify how significant the difference between these biasing relations is,
we proceed as in section \ref{sec:example}. We draw $10^4$ Monte Carlo realizations from 
the redshift distributions of the physically associated cluster pairs
shown in the bottom right panel of figure \ref{fig:bias_comp}.
We compute the variable $\chi^2_{i\vert j}$ (equation \ref{eq:chi2}), where
now $i$ and $j$ stay for our fiducial (equation \ref{eq:bias}), scale independent
(equation \ref{eq:bias_no-r}) or time independent (equation \ref{eq:bias_no-z})
biasing relations.

Our results are shown in figure \ref{fig:bias_chi2}. 
The difference between our fiducial model and the scale independent model
generates large $\chi^2$ values when compared to each other (top panels).
The difference between our fiducial
model and the time independent model, although substantially smaller, is still
well measurable (bottom panels).

\section{Discussion and conclusion}\label{sec:concl}

One of the main purposes of SZ cluster surveys is to constrain the cosmological
model with the cluster redshift distribution. 
In fact, the redshift distribution of SZ selected clusters depends on the
density parameter $\Omega_0$, the mass power spectrum, and the
physical properties of the ICM.  Therefore one hopes that catalogues of
SZ clusters can constrain the cosmological models despite the
degeneracies between some parameters such as the normalization of the power
spectrum $\sigma_8$ and $\Omega_0$ (\citealt{barbosa96};
\citealt{kneissl01}; \citealt*{holder01}; \citealt*{benson02}).

However, for the very same reason why SZ cluster surveys are
so attractive, their catalogues will not have any accurate information on the 
cluster distances, although the morphology of the cluster can provide a rough estimate
of its redshift \citep{diego02}.  
The redshift distribution of clusters
can be determined with follow up optical observations, CO line measurement in galaxy
cluster members, or high resolution X-ray spectroscopy (e.g. \citealt{bleeker02}).
The number of objects in SZ cluster surveys depends on the sky coverage,
instrumental sensitivity, and observational bands.  In an all sky
surveys, this number can be of order several
thousands.  Of course, only the redshifts of a subsample of these
clusters is sufficient to provide a robust estimate of the redshift
distribution. How should one choose this cluster subsample? 
We suggest to measure the redshift of clusters in close pairs.
Because the probability that clusters close on the sky are
also close along the line of sight is substantial, this strategy
will provide, at the same time, the redshift distribution of clusters and a catalogue of
superclusters. 

To show that the compilation
of a catalogue of superclusters from SZ cluster surveys is indeed feasible, 
we have computed the angular correlation function of SZ selected clusters 
for two cosmological models. We have then 
computed how many cluster pairs, close on the sky, are also close
along the line of sight.  For example, for clusters with specific 
flux difference larger than $200$ mJy at $143$ GHz in a $\Lambda$CDM
model, $70\%$ of the pairs with angular separations smaller
than $20$ arcmin are physical associations.
Because an SZ supercluster catalogue will contain superclusters at any redshifts thanks to
the redshift independence of the SZ effect, the catalogue
will be of great relevance 
for follow up studies of the
evolution of galaxy and cosmic structures in high density regions.
Moreover, from the redshift distribution of physically associated cluster pairs, we can extract
information on the biasing relation between the cluster and the mass distribution.

To compute the number of detectable clusters, we 
have used the point source approximation, where the survey
can detect the total cluster radio flux, but it cannot resolve the cluster
structure. This approximation is thus independent
of the cluster profile. It is realistic for poor resolution surveys (for example 
the Planck surveyor will perform a survey with a beam FWHM of 8 arcmin at 143 GHz),
but of course unrealistic for high resolution bolometric or interferometric
ground-based surveys. However, resolving the clusters has the net effect of increasing
the mass threshold and thus decreasing
the number of detected sources \citep{bart00}. In other words, resolving the
cluster structure is similar to 
considering our results for shallower surveys: the cluster
minimum mass is larger and the fraction of physically associated pairs increases.

\section*{ACKNOWLEDGMENTS}  
We thank Lauro Moscardini for fruitful discussions.  The Hubble Volume
$N$-body simulations used in this paper were carried out by the Virgo
Supercomputing Consortium using computers based at the Computing
Centre of the Max-Planck Society in Garching and at the Edinburgh
parallel Computing Centre.  The data are publicly available at
http://www.mpa-garching.mpg.de/NumCos.  This research was partially
supported by the NATO Collaborative Linkage Grant PST.CLG.976902, 
the Italian MIUR grant COFIN2001028932 ``Clusters and
groups of galaxies: the interplay of dark and baryonic matter'',
the EC RTN network ''The Physics
of the Intergalactic Medium'',
and a grant from the G.I.F., the German Israeli
Foundation for Scientific Research and Development.
AN is supported by a grant from the Israeli Science Foundation
and the Technion Fund for the Promotion of Research.
We acknowledge use of the abstract server at the NASA's Astrophysics
Data System and the archive of the astro-ph preprint server.

\end{document}